\documentclass[pra,amsmath,aps,twocolumn]{revtex4} 

\usepackage{graphicx}
\newcommand{\da}{^{\dagger}}

\renewcommand{\r}{{\bf r}}
\newcommand{\br}{({\bf r})}

\newcommand{\bra}{\langle}
\newcommand{\ket}{\rangle}
\newcommand{\hsp}{\hat{H}^{\rm sp}}
\newcommand{\mhsp}[2]{\bra #1 |\hsp| #2\ket}
\newcommand{\mvs}[4]{\bra #1 #2 |V^{\rm s}| #3 #4\ket}
\newcommand{\mv}[4]{\bra #1 #2 |V| #3 #4\ket}
\newcommand{\nn}{\nonumber}
\newcommand{\half}{\frac{1}{2}}

\newcommand{\angles}[1]{\langle#1\rangle}

\newcommand{\tbec}{T_{\rm c}}

\newcommand{\bal}{\begin{align}}
\newcommand{\eal}{\end{align}}

\newcommand{\be}{\begin{equation}}
\newcommand{\ee}{\end{equation}}
\newcommand{\ba}{\begin{eqnarray}}
\newcommand{\ea}{\end{eqnarray}}
\newcommand{\bastar}{\begin{eqnarray*}}
\newcommand{\eastar}{\end{eqnarray*}}

\begin{document}

\title{Collapse and revival of excitations in Bose-Einstein condensates}
\author{M.~M\"ott\"onen, S.~M.~M.~Virtanen and M.~M.~Salomaa}
\affiliation{Materials Physics Laboratory, 
Helsinki University of Technology\\
P.~O.~Box 2200 (Technical Physics), FIN-02015 HUT, Finland}

\date{\today}

\begin{abstract}
We study the energies and decay of elementary excitations
in weakly interacting Bose-Einstein condensates within 
a finite temperature gapless second order theory. The energy shifts for the
high-lying collective modes turn out to be systematically negative compared with the 
Hartree-Fock-Bogoliubov-Popov approximation and the decay of the low-lying
modes are found to exhibit collapse and revival effects. 
In addition, perturbation theory is used to qualitatively explain the experimentally 
observed Beliaev decay process of the scissors mode.
\end{abstract}

\pacs{PACS number(s): 03.75.Kk, 67.40.Db, 05.30.Jp}

\maketitle

\section{Introduction}
The partially Bose-Einstein condensed trapped atomic gases provide
an excellent testbench for developing finite temperature quantum
theories. These weakly interacting systems can be 
modelled from first principles, and the experiments yield accurate 
and detailed information for comparison.
Especially, the energies and decay rates of low-energy collective 
excitations have been measured at different temperatures
and the results provide stringent tests for theoretical models. 

For dilute condensates at temperatures much lower than the condensation temperature $\tbec$, the Bogoliubov approximation consisting of the Gross-Pitaevskii (GP) equation for the condensate wave function and the Bogoliubov equations for the quasiparticle excitations 
has proven to be accurate in describing the collective modes of the system. 
For higher temperatures one has to take into account the effects of the
thermal gas component. Developing a theory that is computationally feasible
and correctly models the system at temperatures approaching $\tbec$ is
a challenging task. The most commonly used finite-temperature theory
is the Hartree-Fock-Bogoliubov-Popov (HFB-Popov) approximation. It 
neglects the dynamics of the thermal gas and the modifications in particle 
correlations induced by the condensate, but predicts quasiparticle
energies in fair agreement with the experiments \cite{popov_vs_exp}.
The energy of the  quadrupole modes having azimuthal 
angular momentum quantum numbers $q_\theta=\pm 2$ deviates from the 
theoretical prediction for temperatures above $0.6\tbec$, but lately this
deviation has been interpreted to mainly arise from improper 
modelling of the time-dependent external potential used in the experiments to
excite the collective modes \cite{morgan_lett}.

In order to take into account the leading order
quasiparticle interactions and the correlations induced by the condensate
in the inhomogeneous case,
several theoretical approaches have been suggested \cite{tosi,bene,fedichev,brunette,reidl,giorgini,morgan_phd,morgan_iop}.
The dynamics of the condensate and the thermal gas has also been studied 
using various kinetic theories \cite{zgn1,zgn2,zgn3,walsseri,proukakki}. 
The second order theory for inhomogeneous, partially condensed gases presented 
in Refs.~\cite{morgan_phd,morgan_iop} uses systematic perturbation theory 
to take into account the interaction terms in the Hamiltonian. Recently,
this theory was extended to take into account the time-dependent 
external perturbation used to drive the system in the experiments, leading
to an agreement with the measured energies and the 
damping rates of the collective modes \cite{morgan_ext,morgan_lett}.

The second order theories are computationally challenging, and
there has been only a few numerical investigations of their predictions
\cite{morgan_lett,second_ord_lett,mizu}. In this paper, we calculate 
the spectral distributions of the quasiparticle energies for a partially condensed Bose-Einstein condensate (BEC),
and compare the quasiparticle energies to the HFB-Popov results as functions of temperature. 
Especially, we analyze the quasiparticle dynamics implied by the
spectral distributions, observing that some collective modes should
exhibit notable collapse and revival effects in trapped condensates.
The possible existence of this phenomenon has been pointed out 
previously in Refs.~\cite{morgan_iop,morgan_phd} (see also Ref.~\cite{pitaevskii}), but, 
however, it has not been studied in detail before. 
The collapse and revival of the excitations 
indicates that the energies and the damping rates alone do not suffice to 
describe the dynamics of these modes, i.e., the commonly used damped 
sinusoidal fit to the experimental data may not be sufficient to describe
the longer term dynamics of some modes. 

The structure of the paper is the following: In Section \ref{sec2}, we
describe the second order theory used in the analysis. Section \ref{sec3} 
is devoted to a discussion of the numerical methods used to calculate the 
excitation spectra and the dynamics of modes. In Section \ref{sec4}, we 
analyze the second order corrections to the excitation spectrum 
as functions of temperature, and in Section \ref{sec5} we study the 
decay of certain modes. Section 
\ref{sec6} consists of discussion and summary of the results.

\section{Second order theory}\label{sec2}
In this section, following Refs.~\cite{morgan_iop,morgan_phd}, we
present the second order formalism for calculating the quasiparticle
spectral distributions for a partially condensed, dilute, trapped
BEC at finite temperatures. The starting point is the usual second 
quantized Hamiltonian for structureless bosons
\be\label{sechamil}
\hat{H}=\sum_{ij}\mhsp{i}{j}\hat{a}\da_i\hat{a}_j+\half \sum_{ijkm}\mv{i}{j}{k}{m}\hat{a}\da_i\hat{a}\da_j\hat{a}_k\hat{a}_m,
\ee 
where the creation and annihilation operators for a particle in state $|i\ket$ are denoted by $\hat{a}\da_i$ and $\hat{a}_i$, respectively. The single particle Hamiltonian is given by the sum of the kinetic energy and the external trapping potential as 
$$
\hsp=-\frac{\hbar^2\nabla^2}{2m}+V_{\rm trap}(\r),
$$
and the dominant $s$-wave scattering at low temperatures can be 
modelled by the effective low energy interaction potential
\be\label{ipot}
V(\r)=\frac{4\pi a \hbar^2\delta(\r)}{m},
\ee
where $a$ is the scattering length and $m$ the atomic mass. 
This effective potential is inapplicable at high energies and leads to 
ultraviolet divergences in the theory which have to be renormalized
in a proper way, see Appendix~\ref{appa}.

We choose to use a canonical ensemble with fixed total number of particles $N$. By defining the bosonic number conserving operators 
$
\hat{\alpha}_i=[(\hat{N}_0+1)^{-1/2}\hat{a}_0]\da\hat{a}_i,
$
where the index $0$ refers to the condensate state and 
$\hat{N}_0=\hat{a}_0\da\hat{a}_0$, one can write the
Hamiltonian (\ref{sechamil}) as
\be\label{hsum}
\hat{H}=\sum_{i=0}^4\hat{H}_i+O(N_0[\hat{\delta}/N_0]^{5/2}),
\ee
where
\ba\label{hs}
\hat{H}_0 \!\!&=&\!\! N_0\left[\mhsp{0}{0}+\half N_0\mvs{0}{0}{0}{0} \right], \\
\hat{H}_1 \!\!&=&\!\! \sqrt{N_0}\sum_{i\neq 0}\!\left[\mhsp{i}{0}\!+\!N_0\mvs{i}{0}{0}{0} \right]\hat{\alpha}_i\da\!+\!{\rm h.c.}, \\
\hat{H}_2 \!\!&=&\!\! \sum_{ij\neq 0}\left[\mhsp{i}{j}-\lambda\delta_{ij}+2N_0\mvs{0}{i}{j}{0} \right]\hat{\alpha}_i\da\hat{\alpha}_j \nn \\
\!\!&&\!\!+\sum_{ij\neq 0}\left[\frac{N_0}{2}\mvs{i}{j}{0}{0}\hat{\alpha}_i\da\hat{\alpha}_j\da +{\rm h.c.}\right]+\lambda\angles{\hat{N}_{\rm ex}}, \\
\hat{H}_3 \!\!&=&\!\! \sum_{ijk\neq 0}\left[\sqrt{N_0}\mvs{i}{j}{k}{0}\hat{\alpha}_i\da\hat{\alpha}_j\da\hat{\alpha}_k+{\rm h.c.} \right],\\
\hat{H}_4 \!\!&=&\!\! \sum_{ijkm\neq 0}\half \mvs{i}{j}{k}{m}\hat{\alpha}_i\da\hat{\alpha}_j\da\hat{\alpha}_k\hat{\alpha}_m,
\ea
and $\hat{\delta}=\hat{N}_{\rm ex}-\angles{\hat{N}_{\rm ex}}$ is the
number fluctuation operator of the noncondensate particles. The symmetrized matrix elements of the two-particle interaction potential $V(\r)$ are defined as
$$
\mvs{i}{j}{k}{m}=\frac{1}{2}[\bra ij|V|km\ket+\bra ji|V|km\ket],
$$
and $\lambda$ as
$$
\lambda = \mhsp{0}{0}+N_0\mvs{0}{0}{0}{0},
$$
where the average number of atoms in the condensate state is given by 
$N_0=N-\angles{\hat{N}_{\rm ex}}$.
Above the averages $\angles{\dots}$ refer to quantum expectation values
and ${\rm h.c.}$ stands for hermitian conjugate.

In the zeroth order approximation, one solves the ground state $|0\ket$ of $\hat{H}_0$ alone, which makes the linear Hamiltonian $\hat{H}_1$ to vanish. The excitations are found in lowest order by diagonalizing $\hat{H}_2$ and the number of the condensed particles $N_0$ has to be tuned such that the total number of particles satisfies $N=N_0+N_{\rm ex}$. 

It is convenient the use an orthonormal single-particle basis 
$\zeta_i(\r)=\bra{\r}|i\ket$ for all $i=0,1,\dots\,$, where 
$\zeta_0(\r)$ is the condensate wave function given by the
Gross-Pitaevskii equation
\be\label{gp}
-\frac{\hbar^2}{2m}\nabla^2\zeta_0+V_{\rm trap}(\r)\zeta_0+N_0U_0|\zeta_0(\r)|^2\zeta_0=\lambda\zeta_0(\r),
\ee
with $U_0=4\pi a\hbar^2/m$. The GP equation is obtained
by minimizing $\angles{\hat{H}_0}$ with respect to $\zeta_0(\r)$.
Diagonalizing $\hat{H}_2$ using the Bogoliubov transformation 
$\hat{\beta}_i=\sum_{j\neq 0}\left[UU_{ij}^*\hat{\alpha}_j-V_{ij}^*\hat{\alpha}_j\da\right]$
results in the Bogoliubov equations
\be\label{bg}
\left(
\begin{array}{cc}
{\mathcal{L}(\r)} & \mathcal{M}(\r) \\
-\mathcal{M}^*(\r) &-\mathcal{L}(\r) 
\end{array}\right)
\left(\begin{array}{c}
u_i(\r)\\
v_i(\r)
\end{array}\right)
=
\epsilon_i
\left(\begin{array}{c}
u_i(\r)\\
v_i(\r)
\end{array}\right),
\ee
where $u_i(\r)=\sum_{j\neq 0}U_{ij}\zeta_j(\r)$ and $v_i(\r)=\sum_{j\neq 0}V_{ij}\zeta_j^*(\r)$ are the quasiparticle amplitudes, $\epsilon_i$ the quasiparticle energies, and operators ${\mathcal{L}}(\r)=\hat{H}^{\rm sp}-\lambda+2N_0U_0|\zeta_0(\r)|^2$ and ${\mathcal{M}}(\r)=N_0U_0\zeta_0^2(\r)$ have been introduced. The quasiparticle amplitudes must satisfy the orthogonality and symmetry relations
\ba
\int{\rm d}\r[u_i(\r)u_j^*(\r)-v_i(r)v_j^*(\r)]&=&\delta_{ij},\nonumber  \\
\int{\rm d}\r[u_i(\r)v_j(\r)-v_i(r)u_j(\r)]&=&0, 
\ea
for the Bogoliubov transformation to be canonical. The quasiparticles must also be orthogonal to the condensate state, i.e., $\int{\rm d}\r\zeta_0^*(\r) u_i(\r)=\int{\rm d}\r\zeta_0(\r) v_i(\r)=0$. The Bogoliubov equations
have the zero-energy solution $\left\{u_0(\r), v_0(\r)\right\}=\left\{\zeta_0(\r), -\zeta_0^*(\r)\right\}$, and
projection to this homogeneous solution should always be subtracted from the
quasiparticle amplitudes.

To calculate the next lowest-order mean fields, i.e., the density of the thermal atoms $\rho(\r)=\sum_{ij\neq 0}\zeta_j^*(\r)\zeta_i(\r)\bra\hat{\alpha}_j\da\hat{\alpha}_i\ket$ and the so-called anomalous average $\kappa(\r)=\sum_{ij\neq 0}\zeta_j(\r)\zeta_i(\r)\bra\hat{\alpha}_j\hat{\alpha}_i\ket$, we express the particle operators in terms of the quasiparticle operators, yielding
\ba
\rho(\r)&=&\sum_{i\neq 0}\left\{[|u_i(\r)|^2+|v_i(\r)|^2]n_i+|v_i(\r)|^2\right\}, \\
\kappa(\r)&=&\sum_{i\neq 0}u_i(\r)v_i^*(\r)(2n_i+1).
\ea
In principle, the quasiparticle populations $n_i=\bra\hat{\beta}_i\da\hat{\beta}_i\ket$ should be calculated from the requirement that the canonical partition function ${\mathcal{Z}}_{\rm c}=\sum_{\{ n_i\}}e^{-\beta E_i(\{ n_i\})}$ minimizes the free energy ${\mathcal{F}}=-k_{\rm B}T\log {\mathcal{Z}}_{\rm c}$. However, to a good approximation \cite{bergeman} one may use the non-interacting quasiparticle gas result $n_i=(z^{-1}e^{\beta\epsilon_i}-1)^{-1}$, where the fugacity is calculated from the relation $z=N_0/(1+N_0)$.

In calculating the perturbative corrections to the zeroth-order theory
corresponding to Eqs.\ \eqref{gp} and \eqref{bg}, it is convenient to first 
calculate the improved condensate wave function 
$\tilde{\zeta}_0(\r)$ from the generalized 
Gross-Pitaevskii (GGP) equation
\ba
\!\!&-&\!\!\frac{\hbar^2}{2m}\nabla^2\tilde{\zeta}_0(\r)+V_{\rm trap}(\r)\tilde{\zeta}_0(\r)+N_0U_0|\tilde{\zeta}_0(\r)|^2\tilde{\zeta}_0(\r) \nonumber \\
\!\!&+&\!\!2U_0\rho(\r)\tilde{\zeta}_0(\r)+U_0\kappa(\r)\tilde{\zeta}_0^*(\r)=\lambda_{\rm g}\tilde{\zeta}_0(\r),\label{ggp}
\ea
which is obtained by minimizing 
$\angles{\hat{H}_0}+\angles{\hat{H}_2}$. 
Expressing the terms in the Hamiltonian as
\be
\hat{H}_i=\hat{H}_i[\zeta_0]+\Delta\hat{H}_i,
\ee
where 
\be
\Delta\hat{H}_i=\hat{H}_i[\tilde{\zeta}_0]-\hat{H}_i[\zeta_0],
\ee
one finds the perturbative Hamiltonian
\be\label{hpert}
\hat{H}_{\rm pert}=\Delta\hat{H}_0+
\Delta\hat{H}_1+\Delta\hat{H}_2+\hat{H}_3+\hat{H}_4,
\ee
where the non-quadratic terms $\hat{H}_3$ and $\hat{H}_4$ are to be calculated using the improved
condensate wave function. Note that our notation for $\Delta \hat{H}_i$
differs somewhat from that in Refs.\ \cite{morgan_iop,morgan_phd}.

The perturbation term $\Delta\hat{H}_0$ is just a real number and can be
easily taken into account. In addition to it, in first order perturbation 
theory only the terms $\Delta\hat{H}_2$ and $\hat{H}_4$ containing
even numbers of quasiparticle operators contribute to the energy shift
\be
E_{\rm pert}(s,1)=\bra s|\hat{H}_{\rm pert}|s\ket,
\ee
where $|s\ket$ is a quasiparticle occupation number eigenstate. In second order
perturbation theory, one can in fact neglect the terms
$\Delta\hat{H}_2$ and $\hat{H}_4$, because it turns out that
their contribution is of the same order as the contribution
of the other terms in third order perturbation theory~\cite{morgan_iop,morgan_phd}. Thus, one
only needs to calculate
\be
E_{\rm pert}(s,2)\approx
\sum_{r\neq s}\frac{|\bra r|\Delta\hat{H}_1+\hat{H}_3|s\ket
|^2}{E_s-E_r}.
\ee

The quasiparticle energies are calculated as total energy changes
in the system when the corresponding quasiparticle occupation number
is increased by one, while the total number of particles is held constant.
This yields the corrected excitation energy 
\be\label{ep}
E_p(z')=\epsilon_p+\Delta E_{4}^p+\Delta E_{\rm shape}^p+\Delta E_{\lambda}^p+\Delta E_{3}^p(z'),
\ee
where the $\Delta$-terms are given in Eqs.~(\ref{e4})-(\ref{e3}) and the complex energy parameter $z'$ should not be mixed with the fugacity. 
Calculating the excitation energies as functions of $z'$ yields the
dynamics of the excitations in the following way: 
The time evolution operator $\hat{U}(t)$ of the system may be written in terms of the Fourier transform of the resolvent operator $\hat{G}(z')=(z'-\hat{H})^{-1}$ as \cite{optic}
\be
\hat{U}(t)=-\frac{\hbar}{\pi}\int_{-\infty}^\infty e^{-i\omega t}{\rm Im}[\hat{G}(\hbar\omega-i0)]d\omega.
\ee
Let us define the projection of the resolvent to state $p$ as $G_p(z)=\bra p|\hat{G}|p\ket$, which may be approximated to second order as
\be\label{resp}
G_p(\omega)=[\hbar\omega-E_p(\hbar\omega)]^{-1}.
\ee
Finally, it is seen that the imaginary part of the projected resolvent $F_p(\omega-i0)={\rm Im}[G_p(\omega-i0)]$ gives the spectral distribution of the mode $p$ and the Fourier transform of $F_p(\omega)$ yields its time dependence. 

The need to calculate quasiparticle energies as functions of $z'$ is
naturally related to the fact that one takes into account quasiparticle
interactions, though only to the lowest order, and the quasiparticle
states are no more energy eigenstates having infinite lifetime. 
In addition, in computing the quasiparticle 
energies $z'$ must have a small imaginary
part acting as a regularizer for the otherwise divergent expressions
for second order energy shifts. One may note that setting $z'=\epsilon_p$
yields the usual Rayleigh-Schr\"odinger perturbation theory, 
while the Brillouin-Wigner perturbation theory corresponds to solving the 
equation $E_p(z')=z'$.

In conclusion, the second order theory may be used to calculate the energies and the dynamics of quasiparticles. First the GP equation (\ref{gp}) is solved together with the Bogoliubov equations (\ref{bg}) for a given total particle number $N=N_0+N_{\rm ex}$. Then the GGP equation (\ref{ggp}) is solved, after which the spectrum $F_p(\omega)$ may be extracted for each excitation $p$ using the energy corrections presented in Eqs.~(\ref{e4})-(\ref{e3}). In addition, one has to take care of proper 
ultraviolet renormalization; The quantities $\kappa(\r)$ and $\Delta E_3^p$ are to be replaced by their renormalized values given in Eqs.~(\ref{kappar}) and (\ref{e3r}) in all calculations.

\section{Numerical methods}\label{sec3}
We consider a pancake-shaped system in a harmonic potential 
$$
V_{\rm trap}(\r)=\half m\omega_x^2x^2+\half m\omega_y^2y^2+\half m\omega_z^2z^2,
$$
where the trapping frequencies are $\omega_r=\omega_x=\omega_y$ and $\omega_z$, with 
$\omega_z\gg\omega_r$. For a sufficiently strong trapping potential in the
$z$-direction, the condensate wave function and the thermodynamically
relevant quasiparticle amplitudes can be approximated to be in
cylindrical coordinates $(r,\theta,z)$ of the factorized
form 
\be\label{z0n}
\zeta_0\br=\zeta_0(r)\sigma(z)e^{i m\theta},
\ee
and
\ba\label{un}
u(\r)&=&u(r)\sigma(z)e^{i(q_{\theta}+m)\theta},\\
v(\r)&=&v(r)\sigma(z)e^{i(-q_{\theta}+m)\theta}\label{vn},
\ea
where $\sigma(z)=e^{-z^2/(2a_{z}^2)}/\sqrt{a_{z}\pi^{1/2}}$ is a Gaussian
profile and $a_i=\sqrt{\hbar/m\omega_i}$ are the harmonic oscillator lengths
of the trap. In the following, we consider only the case $m=0$ of an 
irrotational condensate. Using Eqs.~(\ref{z0n}), (\ref{un}) and 
(\ref{vn}), the Gross-Pitaevskii and Bogoliubov 
equations reduce to equations of the radial coordinate only. 
The $z$-dependence of the GP, the GGP and the Bogoliubov equations is 
reduced explicitly by multiplying them with $\sigma(z)$ and integrating 
over $z$. This results in equations similar to the original ones except 
that the interaction strength $U_0=4\pi\hbar^2a/m$ is replaced with its 
quasi two-dimensional version $U_0^{\rm 2D}=2\sqrt{2\pi}\hbar\omega_za_za$ 
and the chemical potential is shifted by $\hbar\omega_z/2$. We use
$a_r$, $1/\omega_r$ and $\hbar\omega_r$ as units of length, time and energy,
respectively. In these units the dimensionless interaction strength 
becomes $U_0^{\rm 2D}=2\sqrt{2\pi}a/a_z$. A peculiarity of the 
reduced equations is that they are independent of the trapping 
frequency $\omega_r$, and hence our results apply for all $\omega_r$, 
provided that $\omega_z\gg\omega_r$.

In numerical calculations, we use spatial discretization and 
finite-difference methods. The ground state solutions of the non-linear 
GP and GGP equations are found by a norm-conserving imaginary time 
integration method based on the Crank-Nicholson scheme. The computation 
speed is enhanced by using a multigrid method, in which the grid is
made gradually denser during the computation. On the other hand, 
in spatial discretization the Bogoliubov eigenvalue 
equation becomes a matrix eigenvalue equation, with the coefficient
matrix having a narrow band. The eigenvalues of this matrix are found 
by implicitly restarted Arnoldi method implemented in the numerical 
library ARPACK. The quasiparticle amplitudes are calculated 
from the Bogoliubov equations 
up to some cut-off energy $E_{\rm cut}$, above which a semiclassical 
approximation \cite{semiclass} is used for calculating their contribution
to mean-field potentials. The use of the semiclassical approximation 
enhances the convergence of the results as a function of 
$E_{\rm cut}$. 

The anomalous average and the energy shifts $\Delta E_3^p$ contain
ultraviolet divergences due to the interaction contact potential
approximation, and they have to be renormalized.
The renormalization scheme presented in the Appendix \ref{appa} shows 
that the divergent part of the anomalous average is proportional to 
$\tilde{\zeta}_0^2(\r)$. Using this information we determine the value 
of the interaction correction $\Delta U_0(E_{\rm cut})$ due to the 
excitations below $E_{\rm cut}$. This coefficient also suffices to
determine the proper renormalization subtraction for the energy shift 
$\Delta E_3^p$. As an approximation to this proper renormalization,
the ultraviolet divergences can also be removed by neglecting the 
zero-temperature parts of the terms $\kappa(\r)$ and $\Delta E_3^p$.
For dilute gases this should be a good approximation \cite{morgan_phd}.
This simpler renormalization method is also computationally 
much faster, since the summations in the energy correction $\Delta E_3^p$ 
converge and many terms may be neglected within computational accuracy. 
In the more accurate renormalization scheme that we have used, 
the summations diverge and 
hence all the terms up to $E_{\rm cut}$ have to be taken into account.

In calculating the spectral distributions 
$F_p(\omega-i\gamma)$, we have to use a finite 
imaginary part $\gamma$ to avoid divergences in Eq.~(\ref{e3}). The value 
of $\gamma$ is estimated for each excitation and temperature separately 
to be small enough for not to affect the mean value of the spectral
distribution nor its Fourier transform in the regime we have presented it.
In a finite system, the spectral
distributions consist of discrete Lorentzian peaks with widths proportional 
to $\gamma$. Thus, when calculating the spectral density, the smaller the regulator $\gamma$ is, the finer
grid for the real part $\omega$ must be 
used. Together with the double 
summation in Eq.~(\ref{e3}), this can increase the computational cost
of the spectrum to be orders of magnitude larger than the cost
needed in solving the excitations from the Bogoliubov equations. 
To make the computation of the spectrum more efficient, the terms 
$A_{ijk}$ and $B_{ijk}$ in Eq.~(\ref{e3}) are calculated only once and
stored in memory. In addition, a comparable speedup is achieved by 
regrouping the summation terms according to their behaviour as
functions of $\omega$: for slowly varying terms, one can use
much sparser $\omega$-discretization.

We note that in the end we take the limit $\gamma\rightarrow0+$, 
as was suggested in Ref.~\cite{morgan_phd}. However, 
in Refs.~\cite{second_ord_lett,morgan_lett,morgan_ext} a finite value 
of the order $\gamma\sim 10^{-2}\omega_{r}$ was used, motivated by the 
finite experimental observation time, and in Ref.~\cite{mizu} the value of $\gamma$ was taken to be $5\times 10^{-3}\omega_{r}$. However, the inclusion of a finite 
$\gamma$ is only one way to model the finite observation time. 
Since the finite value of the regulator $\gamma$ has also an unphysical 
effect of shifting the excitation energies, it might not be the best 
way to model the restricted observation time. 

\section{Excitation spectra}\label{sec4}

\begin{figure}
\includegraphics{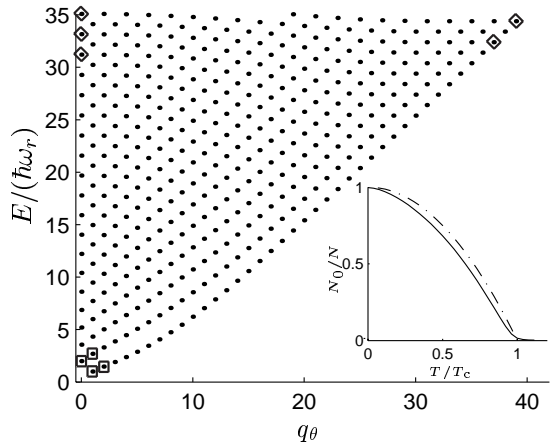}
\caption{\label{fig1} Excitation energies of the lowest 
part of the spectrum at zero temperature. The energies of excitations marked 
with squares and diamonds are presented as functions of temperature in 
Figs.~\ref{fig2} and~\ref{fig3}, respectively. The inset shows 
the condensate fraction as function of temperature (solid line),
and the exact result $1-(T/\tbec)^2$ for the non-interacting system 
(dashdot line).}
\end{figure}

In this section, we present and analyze the results for the mean energies of
excitations as functions of temperature. We model a pancake-shaped
cloud consisting of $N=2000$  $^{23}$Na atoms trapped in the tight
direction  with the trapping frequency $\omega_z=2\pi\times 350~{\rm Hz}$. 
As pointed out in Sec.~\ref{sec3}, the radial trapping frequency 
$\omega_r=\omega_x=\omega_y$ may be chosen freely with only the constraint 
$\omega_z\gg\omega_r$. These parameters are chosen for convenience
to coincide with the ones used in Ref.~\cite{morgan_2d}, in which the 
excitation energies were calculated from the self-consistent HFB-Popov theory.
Previously, energies of a few modes for spherically symmetric systems
\cite{second_ord_lett} and condensates containing a vortex line 
\cite{mizu} have been computed within this second order theory.

\begin{figure}
\includegraphics{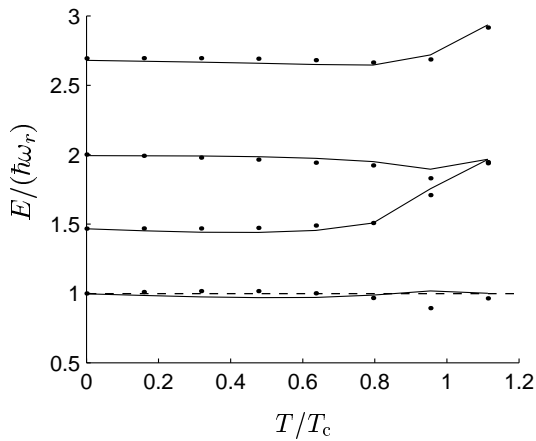}
\caption{\label{fig2}Temperature dependence of the mean energies of the 
modes marked with squares in Fig.~\ref{fig1}. Dots correspond to
the second order theory, and solid lines to the HFB-Popov theory. The
dashed line indicates the exact energy $\hbar\omega_r$ of the 
Kohn modes.}
\end{figure}

Figure~\ref{fig1} shows the zeroth order energies of the lowest energy
excitations at zero temperature. Within the accuracy of 
the figure, the zeroth order Bogoliubov energies 
coincide with those of the full second order theory. Since the 
condensate is irrotational, the spectrum is symmetric with respect to
inversion $q_\theta\rightarrow -q_\theta$, and thus only excitations 
with non-negative angular momenta are shown. The inset 
presents the condensate fraction 
as a function of the temperature (solid line) compared with the 
non-interacting gas result $N_0^{'}(T)/N=1-(T/\tbec)^2$ (dashdot line). 
We identify the condensation temperature $\tbec$ as the point where
the condensate fraction obtains its maximum 
second derivative with respect to the temperature. The theory is 
probably not reliable above or in the vicinity $\tbec$, although we present its predictions
also in this regime.

\begin{figure}
\includegraphics{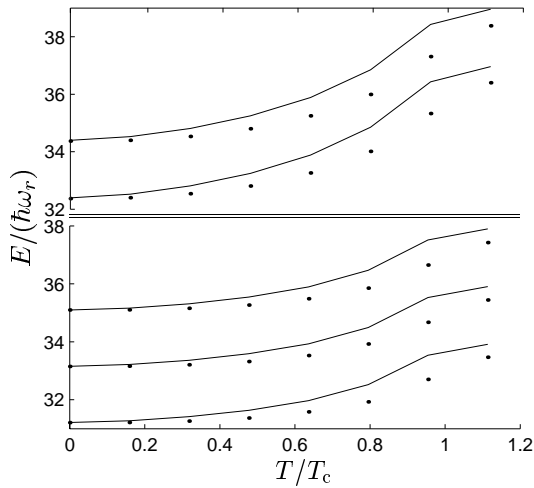}
\caption{\label{fig3}Temperature dependence of the mean energies of the 
modes marked with diamonds in Fig.~\ref{fig1} within second order theory
(dots) compared with the HFB-Popov results (solid lines). 
The energies in the lower block correspond to excitations with 
vanishing angular momentum and the upper block to excitations with 
$q_\theta\approx 40$.}
\end{figure}

The energies of the modes marked with squares and diamonds in 
Fig.~\ref{fig1} are presented as functions of temperature in Figs.~\ref{fig2} 
and~\ref{fig3}, respectively. The mean values of the spectral 
distributions of the excitations within the second order theory 
are shown, in addition to the corresponding HFB-Popov results
that were obtained by neglecting the anomalous average, the second 
order corrections and the terms in the second line of Eq.~(\ref{e4}). 
For the low-lying modes, some of the second order energies are observed 
to cross the Popov results as seen in Fig.~\ref{fig2}. 
On the other hand, Fig.~\ref{fig3} shows that the second order 
theory for the higher lying modes yields systematically lower 
excitation energies than the Popov theory. 
The Popov results seen in Fig.~\ref{fig2} 
are consistent with the energies calculated in Ref.~\cite{morgan_2d}.

In Fig.~\ref{fig2}, the dashed line corresponds to the energy $\hbar\omega_r$
of the exact center of mass oscillation modes, the Kohn modes. 
According to the generalization~\cite{kohn_trap_int,kohn_op} of Kohn's theorem~\cite{kohn}, a system of harmonically trapped interacting particles in 
any eigenstate of the Hamiltonian has an eigenstate with the amount 
$\hbar\omega_i$ higher energy, i.e., the exact diagonalization of the Hamiltonian should yield a spectrum with the eigenenergy $\hbar\omega_i$. The Bogoliubov theory, in which the thermal gas component is neglected, implies Kohn modes to have this exact 
energy. In the higher order theories, the dynamics of the thermal gas 
and its interaction with the condensate have to be taken into 
account accurately to obtain results in agreement with the Kohn theorem.
Figure~\ref{fig2} shows that within the second order theory the energy 
of the Kohn mode is very close to $\hbar\omega_r$ for 
temperatures $T<0.8\tbec$. Taking into account perturbation
theory terms beyond the second order should yield
energies even closer to the exact result.

The lowest mode with vanishing angular momentum is the 
breathing mode corresponding to uniform scaling oscillations of the
condensate. In the case of a two-dimensional harmonically 
trapped gas interacting via the 
contact potential, it has been shown using 
the scaling symmetry of the Hamiltonian that there exists a state 
that has energy $2\hbar\omega_r$ in excess to the ground 
state~\cite{breath}. This excitation is identified with the 
breathing mode. The Bogoliubov theory yields exactly the 
energy $2\hbar\omega_r$ for the breathing mode, while the 
Popov and the second order theories do not, as can be seen in Fig.~\ref{fig2}. 
Since the interaction potential has to be renormalized and hence 
deviates from that used in Ref.~\cite{breath} for modes 
with high energy, the applicability of the exact result is 
somewhat questionable at high temperatures, where the physics 
is not determined by the low-lying modes alone.
It is shown in Fig.~\ref{fig2} that the energy of the breathing 
mode is lower than $2\hbar\omega_r$ and the deviation from 
$2\hbar\omega_r$ increases with the temperature.

\section{Decay of the excitations}\label{sec5}

In experiments and theoretical studies, the decay of an excitation is 
commonly characterized only by the damping rate related 
to the exponential decay of the oscillation amplitude. 
For infinite systems, the excitation spectrum is continuous and
the spectral distributions $F_p(\omega)$ of the excitations are
Lorentzian peaks, implying indeed an exponential decay of the mode 
oscillations. The mean value of the 
Lorentzian gives the mode frequency and its width the damping rate.
However, for trapped, finite systems the spectrum is
discrete and the spectral distributions generally have
more complicated forms. Especially, the dynamics implied by 
these distributions can be more complicated than just the
simple exponential decay. From the computed spectral distributions
of the oscillations, we have studied the validity of the 
exponential decay approximation for the finite system under question. 

\begin{figure}
\includegraphics{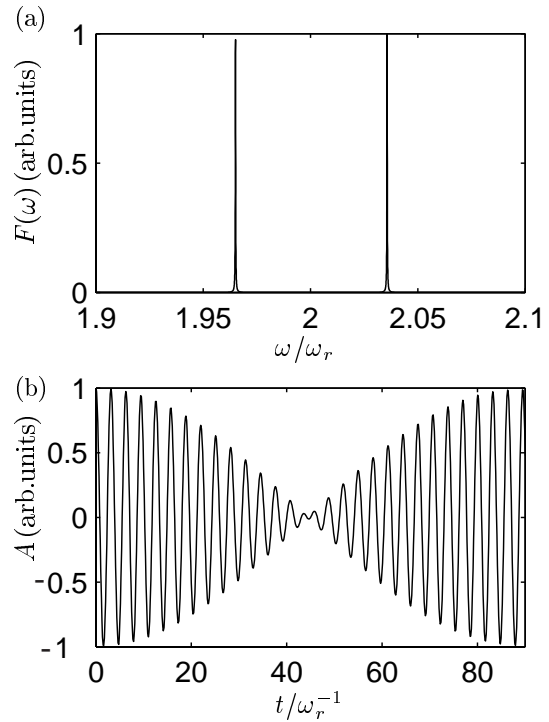}
\caption{\label{fig4} (a) Spectral distribution $F(\omega)$ of the 
breathing mode (the lowest mode with $q_\theta=0$) at zero temperature
and (b) its Fourier transform.}
\end{figure}

In fact, the computed spectral distributions consist of
discrete peaks, as noted also previously \cite{second_ord_lett,mizu},
and the form of the distributions is typically far from a simple
Lorentzian profile. This seems to imply the excitation amplitudes to have a 
complicated modulation in time. The extreme case of this modulation 
is the collapse and revival of the corresponding excitation amplitude. 
This phenomenon can be seen for the breathing mode at zero 
temperature. Figure~\ref{fig4}(a) displays the computed spectral distribution
for the breathing mode, which consists of two large, well separated peaks.
This implies strong beating behaviour in the mode amplitude, seen in
Fig.~\ref{fig4}(b), in which the amplitude of 
the oscillation collapses in time $t=45/\omega_r$, 
but revives as time elapses. In fact, the amplitude 
has a beating behaviour with a base frequency given by  the 
mean value of the peaks in Fig.~\ref{fig4}(a), and a beating 
frequency inversely proportional to the distance between the peaks. 

The two peaks in Fig.~\ref{fig4}(a) are due to  
a Beliaev process \cite{beliaev_process} resonance, in 
which a breathing mode quasiparticle 
with Bogoliubov energy $2\hbar\omega_r$ decays into two 
Kohn mode quanta with opposite angular momenta and energy 
$\hbar\omega_r$. Owing to the temperature independent terms 
in Eq.~(\ref{e3}), the Beliaev process may take place even at 
zero temperature. For the simpler, approximate renormalization scheme in which the 
temperature independent terms 
in the anomalous average and the energy correction $\Delta E_3$ 
are completely neglected, this Beliaev process cannot occur at zero 
temperature, and the spectral distribution consists of only a 
single peak. For the Kohn mode, there are no modes into 
which it could decay via Beliaev processes and 
hence the oscillation amplitude of the center of mass is constant 
in time at zero temperature. At finite temperatures, the resonant 
Landau process, the inverse of the Beliaev 
process, splits the spectral distribution of the Kohn modes.

The non-trivial spectral distributions of the breathing mode and the 
Kohn modes seem to contradict the results for the exact energies
of these modes discussed above. This is partly due to the accidental 
strong resonant Beliaev and Landau processes which probably weaken the 
accuracy of perturbation theory. If the 
processes, in which the quasiparticles decay into Kohn or 
breathing modes, are neglected by hand as in
Ref.~\cite{pitaevskii_se_dampping_paperi}, the effects 
of the resonances are removed and the spectral distributions 
of these modes become narrower. However, it is not evident that this procedure 
yields more accurate mean energies. In conclusion, a pure 
collapse and revival of the breathing mode is probably only an 
artefact of the second order theory---it would be interesting to
investigate whether higher order calculations would improve the situation
in this respect.

\begin{figure}
\includegraphics{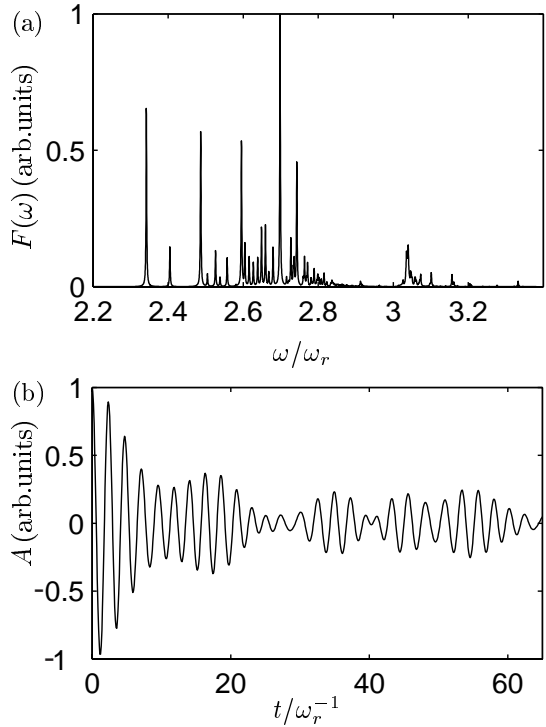}
\caption{\label{fig5}(a) Spectral distribution $F(\omega)$ of the second 
lowest mode with $q_\theta=1$ and (b) its Fourier transform at 
the temperature $T=0.64\tbec$.}
\end{figure}

The spectral distribution and the dynamics of the second lowest mode 
with $q_\theta=1$ at the temperature $T=0.64\tbec$ is shown in 
Fig.~\ref{fig5}. The distribution is obviously far from Lorentzian
form, consisting of several asymmetrically separated peaks, 
and the dynamics is more complicated than the zero temperature 
result for the breathing mode presented in Fig.~\ref{fig4}(b). 
The collapse and revival behaviour is clearly seen, 
although it is weaker than in Fig.~{\ref{fig4}(b)}. 
Our calculations have also showed collapse and ravival of many other elementary excitations.

In Ref.~\cite{oxf_exc3}, the Beliaev decay has been reported to be
observed in  the case of the scissors mode which corresponds 
to scissors-like density fluctuation of the condensate. 
In the experiment, the trapping frequency ratio $\omega_z/\omega_x$ 
was adjusted such that the energy $E_{\rm xz}$ of the scissors mode in  
the $xz$-plane was twice the energy $E_{\rm xy}$ in the $xy$-plane. 
The amplitude of the oscillation is shown in 
Fig.~3(c) of Ref.~\cite{oxf_exc3}, and it was observed that the
amplitude of the mode decreases and increases in time. It 
was also observed that the strength of this phenomenon was peaked 
into the position of the Beliaev resonance as a function of
the trap asymmetry ratio $\omega_x/\omega_y$, implying that
the Beliaev process between these modes is responsible for this effect.

Within the second order formalism, one can interpret this
phenomenon in the following way: Provided that the 
numerator in the term corresponding to the Beliaev decay 
of the scissors mode does not vanish in Eq.~(\ref{e3}) and 
other processes are not important, the second order theory 
yields a spectral distribution proportional to 
$$F_{\rm sc}(\omega)={\rm Im}[1/(\hbar\omega-0i-E_{\rm xz}+A/(\hbar\omega-0i-E_{\rm xz}+E_{\rm ofr}))],$$
where $A$ is the amplitude of the second order correction and $E_{\rm ofr}$ 
is the energy indicating how much the Beliaev process is off-resonant. 
If $E_{\rm ofr}=0$ the distribution $F_{\rm sc}(\omega)$ 
consists of two peaks whose distance is determined by $A$, and the 
dynamics of the mode corresponds a pure collapse and revival 
as shown in Fig.~{\ref{fig4}} in the case of the breathing 
mode. With increasing $E_{\rm ofr}$, one of the two 
peaks becomes smaller and the peaks are shifted in a such 
way that their mean value is kept at $E_{\rm xz}$. This 
corresponds to an oscillation in which the amplitude does 
not vanish completely at any moment, and ultimately when $E_{\rm ofr}\to\infty$
it remains constant, which qualitatively explains the observation 
in the experiments. It is possible that due the parity of the 
scissors modes the second order amplitude vanishes. However, the 
higher order corrections may still have non-vanishing 
amplitudes, resulting in qualitatively same kind of effect.

\section{Conclusions}\label{sec6}
We used the gapless second order theory developed in 
Refs.~\cite{morgan_phd,morgan_iop} to calculate the
excitation energies and dynamics of the collective 
excitations for a partially condensed, harmonically trapped
quasi 2D bosonic gas. The results satisfy
the Kohn theorem quite accurately for temperatures $T<0.8\tbec$. 
The energies of the HFB-Popov and the second order theory crossed 
as functions of temperature for some of the low-lying modes, 
while the second order theory systematically 
yields smaller energies for the higher lying modes. The first 
experimental observations of the Beliaev damping were discussed
within the second order theory and it was found 
that this theory qualitatively accounts for the observations.

The computed spectral densities also imply collapse and revival of many elementary excitations.
The zero temperature spectral distribution of the breathing mode 
is characterized by two large, well-separated peaks and the 
oscillation amplitude consequently displays strong collapse and
revival behaviour within the second order theory. 
This is due to the resonant Beliaev process, in 
which one quasiparticle in the breathing mode decays into two 
quasiparticles in the Kohn modes with opposite angular momenta.
The result seems to contradict exact analytical results
for the breathing mode energy~\cite{breath},
and is probably due to weak convergence of the 
perturbation theory for this mode.
The calculations can in principle be extended to higher order, but
this soon results in overwhelming computational difficulties.
It would also be interesting to upgrade the calculations 
to be self-consistent~\cite{morgan_phd,morgan_iop}, such that
the perturbative energy corrections are inserted into the 
eigenvalue equations, which are then solved iteratively.

\begin{acknowledgments}
The Academy of Finland is appreciated for financial support through a Research Grant in Theoretical Materials Physics (no 201710) and 
CSC-Scientific Computing Ltd (Espoo, Finland) for computational 
resources. M.~M\"ott\"onen acknowledges The Foundation of Technology (Helsinki, 
Finland) and The Finnish Cultural Foundation for financial support. J.~J.~Vartiainen is appreciated for 
discussions. 
\end{acknowledgments}

\appendix*
\section{Ultraviolet renormalization}\label{appa}
Since the low-energy effective
contact potential approximation for the interactions between 
the particles is not valid at high energies, the bare anomalous average 
$\kappa(\r)$ is ultraviolet divergent. To remove this divergence
in a proper way, following Ref.~\cite{morgan_ext}, the interaction 
strength $U_0$ in 
GGP equation (\ref{ggp}) must be replaced by $U_0+\Delta U_0$, where
\be\label{du0}
\Delta U_0=U_0^2\int\frac{{\rm d{\bf k}}}{(2\pi)^3}\frac{m}{\hbar^2k^2}.
\ee
The term $\Delta U_0$ is divergent, and cancels the divergence of the 
anomalous average. Combining the divergent 
terms, one obtains the renormalized anomalous average
\be\label{kappar}
\kappa^{\rm R}(\r)=\kappa(\r)+N_0\frac{\Delta U_0}{U_0}\tilde{\zeta}_0^2(\r),
\ee
which is finite. The GGP equation is now properly renormalized provided 
that the anomalous average is replaced by the renormalized one.

The perturbation Hamiltonian given in Eq.~(\ref{hpert}) is used to 
calculate the total energy of the system $E(N_0,n_1,n_2,\dots)$ up 
to second order in perturbation theory for the given quasiparticle 
distribution $\{n_i\}$. The excitation energies are
obtained as energy differences 
$E_p=E(N_0-\Delta N_p,n_1,n_2,\dots,n_p+1,\dots)-E(N_0,n_1,n_2,\dots)$, 
where $\Delta N_p=\int{\rm d}\r[|u_i(\r)|^2+|v_i(\r)|^2]$ is the 
amount of particles transferred to the mode $p$. The energy
correction terms appearing in Eq.~(\ref{ep}) are given by
\begin{align}
\Delta E_{4}^p &= U_0\int{\rm d}\r \Big[2\rho(\r)\Delta\rho_p(\r)\nonumber \\
&+{\rm Re}[{\kappa^*(\r)\Delta\kappa(\r)}]-\frac{n_p+1}{2}\big\{2\Delta\rho_p^2(\r)\nonumber \\
&+{\rm Re}\left[\Delta\kappa^*(\r)\Delta\kappa(\r)\right]\big\}\Big],\label{e4} \\
\Delta E_{\rm shape}^p &= N_0U_0\int{\rm d}\r \nonumber \\
&\times{\rm Re}\left\{\left[\tilde{\zeta}_0^2(\r)-\zeta_0^2(\r)\right]\Delta\kappa_p(\r)\right\} \nonumber \\
&+ 2N_0U_0\int{\rm d}\r\left[\tilde{\zeta}_0^2(\r)-\zeta_0^2(\r)\right]\Delta\rho_p(\r) \nonumber \\
&+ 2N_0U_0\int{\rm d}\r\,\zeta_0^3(\r)\left[u_p(\r)+v_p(\r)\right] \nonumber \\
&\times\int{\rm d}\r\left[\zeta_0(\r)-\tilde{\zeta}_0(\r)\right]\left[u_p(\r)+v_p(\r)\right],\label{es}
\end{align}
\begin{align}
\Delta E_{\lambda}^p &= (\lambda-\lambda_{\rm g})\int{\rm d}\r\Delta\rho_p(\r), \label{el}\\
\Delta E_{3}^p(z') &= -\sum_{ji\neq 0}\,\!^{'}\bigg[\frac{2|A_{pij}|^2}{z'+\epsilon_i+\epsilon_j} \nonumber \\
&+\frac{2|B_{pij}|^2}{z'-\epsilon_i-\epsilon_j}\bigg](1+n_i+n_j) \nonumber \\
&+\sum_{ij\neq 0}\,\!^{'}\left[\frac{4|B_{ijp}|^2}{z'-\epsilon_i+\epsilon_j}(n_i-n_j)\right]\nonumber \\
&+\sum_{i\neq 0}\,\!^{'}\Big[\frac{4|B_{pip}|^2}{\epsilon_i}(2n_p+1) \nonumber \\
& -2n_p\frac{|B_{ppi}|^2}{2z'-\epsilon_i}-2n_p\frac{|A_{ipp}|^2}{2z'+\epsilon_i}\Big]\label{e3},
\end{align}
where in the primed summations one excludes the terms in which 
all the summation indices are equal to $p$. These diagonal terms are negligible in the current calculations. The contribution to the
density of the thermal gas and the anomalous average due to the mode $p$ are
defined as $\Delta\rho_p({\bf r})=|v_p({\bf r})|^2+|u_p({\bf r})|^2$ and
$\Delta\kappa_p({\bf r})=u_p({\bf r})v_p^*({\bf r})$. Moreover, we have replaced the energy $\epsilon_p$ with a complex variable $z'$. 
The second order matrix elements are written as
\bal
A_{ijk}&=\sqrt{N_0}U_0\int{\rm d}\r\Big\{\zeta_0(\r)\big[v_i\br u_j\br u_k\br\nonumber \\
&+u_i\br v_j\br u_k\br+u_i\br u_j\br v_k\br\big]\nonumber \\
&+\zeta_0^*(\r)\big[u_i\br v_j\br v_k\br \nonumber \\
&+v_i\br u_j\br v_k\br+v_i\br v_j\br u_k\br\big]  \Big\}, \\
B_{ijk}&=\sqrt{N_0}U_0\int{\rm d}\r\Big\{\zeta_0^*\br\big[u_i^*\br u_j\br u_k\br \nonumber \\
&+v_i^*\br v_j\br u_k\br+v_i^*\br u_j\br v_k\br\big]\nonumber \\
&+\zeta_0(\r)\big[u_i^*\br u_j\br v_k\br \nonumber \\ 
&+u_i^*\br v_j\br u_k\br+v_i^*\br v_j\br v_k\br\big]\Big\}.
\end{align}
The energy correction $\Delta E_3^p$ in Eq.~(\ref{e3}) is ultraviolet 
divergent. However, the renormalization of the anomalous average 
implies that the bare second order correction $E^p_3(z)$ is to be
replaced by the renormalized one
\ba
\Delta E_{3{\rm R}}^p(z')\!\!&=&\!\!E^p_3(z')+2N_0\Delta U_0 \nonumber \\ \!\!&\times &\!\!\int{\rm d}\r|\zeta_0(\r)|^2(|u_p(\r)|^2\!+|v_p(\r)|^2).\label{e3r}
\ea

\end{document}